\documentclass[epj]{svjour}

\usepackage{lineno,hyperref}

\usepackage{amsmath}
\usepackage{graphicx,epsfig,wrapfig}
\usepackage[caption=false]{subfig}
\usepackage{mathrsfs}
\usepackage{makeidx}
\usepackage{amsfonts}
\usepackage{amssymb}
\usepackage{amsmath}
\usepackage{mathtools}
\usepackage{fixmath}
\usepackage{amsbsy}
\usepackage{mathrsfs}
\usepackage{slashed}
\usepackage{bm}
\usepackage{multirow}
\usepackage{booktabs}
\usepackage{adjustbox}

\usepackage{color}
\usepackage[normalem]{ulem}

\definecolor{myGreen}{rgb}{0.2,0.72,0.2}
\renewcommand\sout{\bgroup \color[rgb]{0.55,0.00,0.99} \ULdepth=-.5ex \ULset}

\usepackage{simplewick}

\usepackage{multirow}
\usepackage{braket}

\newcommand{\ta}{\left(}
\newcommand{\qa}{\left[}

\newcommand{\tc}{\right)}
\newcommand{\qc}{\right]}

\newcommand{\rDer}[1]{\overset{\rightarrow}{#1}\phantom{\,}}
\newcommand{\lDer}[1]{\overset{\leftarrow}{#1}\phantom{\,}}
\newcommand{\lrDer}[1]{\overset{\leftrightarrow}{#1}\phantom{\,}}

\newcommand{\ds}[1]{\slashed{#1}}

\newcommand{\MS}{{\overline{\text{MS}}}}

\newcommand{\e}{\epsilon}

\renewcommand{\[}{\begin{equation}}
\renewcommand{\]}{\end{equation}}
\allowdisplaybreaks

\modulolinenumbers[5]
\begin{document}
\title{Energy-momentum tensor in the scalar diquark model}
\author{Arturo Amor-Quiroz\inst{1}\inst{2} \and William Focillon\inst{2} \and C\'edric Lorc\'e\inst{2} \and Simone Rodini\inst{2}}
\institute{SoDa, INRIA Saclay, Palaiseau, France \and CPHT, CNRS, Ecole polytechnique, Institut Polytechnique de Paris, 91120 Palaiseau, France}

%\affiliation{SoDa, INRIA Saclay, Palaiseau, France}
%\affiliation{CPHT, CNRS, Ecole polytechnique, Institut Polytechnique de Paris, 91120 Palaiseau, France}
%\author{William Focillon}
%\affiliation{CPHT, CNRS, Ecole polytechnique, Institut Polytechnique de Paris, 91120 Palaiseau, France}
%\author{C\'edric Lorc\'e}
%\affiliation{CPHT, CNRS, Ecole polytechnique, Institut Polytechnique de Paris, 91120 Palaiseau, France}
%\author{Simone Rodini}
%\email{simone.rodini@polytechnique.edu}
%\affiliation{CPHT, CNRS, Ecole polytechnique, Institut Polytechnique de Paris, 91120 Palaiseau, France}
%\author{First author\inst{1} \and Second author\inst{2}% etc

%
%\institute{Insert the first address here \and the second here}
%
\date{}
\abstract{
We study in detail the energy-momentum tensor of the scalar diquark model at the one-loop level using two different regularization methods. We extract the perturbative expressions for the gravitational form factors and we check explicitly that all the Poincar\'e sum rules are satisfied. We confirm in particular that, like in quantum electrodynamics, the symmetric energy-momentum tensor is finite. Finally, we discuss the results for the trace of the energy-momentum tensor and its relation to the mass of the system.
\PACS{
     } % end of PACS codes
} %end of abstract
\maketitle

\section{Introduction}
Recent years have seen a rich discussion on the energy-momentum tensor (EMT) in QCD. Specific focus has been put on the decomposition of the proton mass in QCD, see e.g.~\cite{Ji:1994av,Ji:1995sv,Lorce:2017xzd,Yang:2018nqn,Hatta:2018sqd,Metz:2020vxd,Ji:2021mtz,Lorce:2021xku,Liu:2021gco}, and the relation with the trace of the EMT, which contains a well-known anomalous contribution. Beside the question of the proton spin decomposition, see e.g.~\cite{Jaffe:1989jz,Ji:1996ek,Leader:2013jra,Wakamatsu:2014zza,Lorce:2017wkb,Ji:2020ena,Lorce:2021gxs}, the EMT is also used in the literature to define the notions of pressure and shear distributions inside hadrons~\cite{Polyakov:2002yz,Polyakov:2018zvc,Lorce:2018egm,Freese:2021czn,Burkert:2018bqq,Burkert:2023wzr,Lorce:2023zzg}. More generally, the EMT is a central object in the physics case of the future electron-ion collider in the US~\cite{AbdulKhalek:2021gbh} and can be used to define characteristic sizes of the proton, as illustrated by the recent extraction~\cite{Duran:2022xag}.

In this work we study in detail the EMT for a spin-$\frac{1}{2}$ particle within the scalar diquark model. Contrary to Ref.~\cite{Chakrabarti:2020kdc} whose aim was to provide some realistic predictions for the proton structure based on the light-front wave function representation, we adopt here a perturbative approach to ensure full Poincar\'e covariance. In this case the scalar diquark model is regarded as a mere toy model where fundamental relations can be tested explicitly, giving more insights into the physics. This study complements recent investigations both on the electron EMT in QED~\cite{Rodini:2020pis,Metz:2021lqv,Freese:2022jlu}, and a dressed quark in QCD \cite{More:2023pcy,More:2021stk}. 

We are going to present the complete one-loop results using two different types of regularization. The first is dimensional regularization, which has been used in similar works for QED and QCD. The second is Pauli-Villars regularization, where the emergence of the anomalous contribution to the trace is conceptually different compared to dimensional regularization. We will highlights the key aspects of the comparison between the two regularization schemes. Our perturbative results will then be used to check explicitly various fundamental sum rules derived from Poincar\'e invariance, to analyze the so-called $D$-term, and to investigate the role of the trace anomaly for the mass of the system. The energy, pressure and shear force distributions in Fourier-conjugate space are addressed in the Appendix and show distinct pathological features characteristic of perturbative computations.  

\section{Scalar diquark model}
Different versions of the scalar diquark model exist. The main differences concern the inclusion of different flavors for the quark field, and the inclusion of electromagnetic and color degrees of freedom for the quark and diquark fields. Since at one-loop level for external proton states all the differences amount to at most a global irrelevant factor, we choose to work with the simplest version of the model. The Lagrangian reads
\[
\begin{split}
\mathcal{L} &= \bar{\Psi} \ta\tfrac{i}{2}\lrDer{\slashed{\partial}}-M\tc \Psi + \bar{q} \ta \tfrac{i}{2}\lrDer{\slashed{\partial}}-m\tc q \\
& + \tfrac{1}{2}\ta \partial^\mu \phi \partial_\mu\phi -m_s^2\phi^2\tc + g\phi \ta \bar{\Psi}q + \bar{q}\Psi\tc,
\end{split}
\label{lagrangian}
\]
where $\Psi$ is the proton field, $q$ the quark field and $\phi$ the diquark field, and $\lrDer{\partial} = \rDer{\partial}-\lDer{\partial}$.

We will use two main regularization schemes. The first one is standard dimensional regularization (DR) with $D=4-2\epsilon$, for which no modification of the Lagrangian is needed. The second one is Paulli-Villars (PV) regularization. Specifically, we introduce a ghost field $c$ for the scalar diquark only, for which the Lagrangian reads
\[
\mathcal{L}_\text{PV} = -\tfrac{1}{2}\ta \partial^\mu c\partial_\mu c-M_\text{PV}^2c^2\tc + gc\ta \bar{\Psi}q + \bar{q}\Psi\tc.
\]
Notice that the kinetic term has opposite sign compared to the normal scalar, but the interaction term is identical.

For convenience, we report the equations of motion (EOM) for the theory: 
\[
\begin{split}
&i\ds{\partial}S = \mathbb{M}S, \qquad \bar{S}i\lDer{\ds{\partial}} = -\bar{S}\mathbb{M}, \\
&  \ta\square +m_s^2\tc \phi = g \bar{S}\sigma_1 S, \qquad  
\ta \square+M_\text{PV}^2\tc c = -g\bar{S}\sigma_1 S
\end{split}
\]
with
\[
S=\begin{pmatrix}
    \Psi \\
    q
\end{pmatrix}, \ \  \sigma_1 = \begin{pmatrix}
    0 &1\\
    1&0
\end{pmatrix}, \ \  \mathbb{M} = \begin{pmatrix}
    M & -g(\phi+c) \\
    -g(\phi+c) & m
\end{pmatrix}.
\]
We then derive the symmetric energy-momentum tensor (EMT) via the variation of the action $\int d^4x\sqrt{-g}\,(\mathcal L+\mathcal L_\text{PV})$ with respect to a general metric, evaluated afterwards for the Minkowski metric $g_{\mu\nu}=\text{diag}(+,-,-,-)$. After application of the EOM, we obtain
\[
\begin{split}
T^{\mu\nu} & = \bar{S}\gamma^{\{\mu}\tfrac{i}{2}\lrDer{\partial}^{\!\!\nu\}} S + \partial^\mu \phi \partial^\nu \phi- \partial^\mu c \partial^\nu c\\
& -\tfrac{1}{2}g^{\mu\nu}\ta \partial_\alpha \phi\partial^\alpha \phi - m_s^2\phi^2\tc+\tfrac{1}{2}g^{\mu\nu}\ta \partial_\alpha c\partial^\alpha c - M_\text{PV}^2c^2\tc ,
\end{split}
\label{EMT_operator}
\]
where $a^{\{\mu}b^{\nu\}} = (a^\mu b^\nu + a^\nu b^\mu)/2$. Evidently, in the case of DR the ghost sector is not present.

We are going to investigate the proton matrix elements of Eq.~\eqref{EMT_operator} up to the one-loop level. For this, we will isolate the gravitational form factors, defined by the parametrization 
\begin{align}
& \langle p',s' | T^{\mu\nu}_i| p,s \rangle  = \bar u'\Bigg[\ta A_i(\Delta^2) + B_i(\Delta^2)\tc\frac{P^{\{\mu}i\sigma^{\nu\}\rho}\Delta_\rho}{2M}  \notag\\
& +A_i(\Delta^2)\,\frac{P^\mu P^\nu}{M} +D_i(\Delta^2)\,\frac{\Delta^\mu\Delta^\nu-g^{\mu\nu}\Delta^2}{4M} \label{GeneralEMTParametrization_offF}\\
& + \bar C_i(\Delta^2)g^{\mu\nu} M \Bigg]u,\notag
\end{align}
where $u\equiv u(p,s)$ is the usual free Dirac spinor, $P=(p'+p)/2$ and $\Delta=p'-p$. Notice that the index $i$ labels three different contributions, namely $i=\Psi$ for the proton operators, $i=q$ for the quark operators and $i=d$ for the diquark operators. In the case of PV regularization, we will merge the diquark and ghost contributions.

\section{Loop results}
\begin{figure}[ht]
\centering
\includegraphics[width=0.5\textwidth]{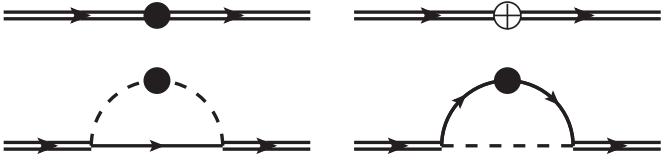}
\caption{Relevant diagrams up to order $g^2$ for the proton matrix elements of the EMT operator. The black dot represents the EMT insertion into the Green's function, the white crossed dot represents the counterterm diagram. Solid, double solid and dashed lines represent the quark, proton and diquark fields, respectively.}
\label{quark_figure}
\end{figure}
In this section we report the main results of this work. First, we present the relevant Lagrangian counterterms in the two chosen regularization schemes. Then we proceed with the results for the EMT insertion in a quark line and in a diquark line, depicted in the second row of Fig.~\ref{quark_figure}.

\subsection{Lagrangian counterterms}
The only necessary counterterms are the ones associated to the proton wave function $Z_\Psi=1+\delta Z_\Psi$ and the proton mass $Z_M=1+\delta Z_M$. We computed them in the on-shell scheme for the proton field. In general, the sum of the counterterm diagrams and the diagrams in which the loops are confined to one of the external legs is independent of the choice of renormalization scheme for the Lagrangian counterterms (individual diagrams though might naturally differ). Since here we are only interested in the total contribution from the proton operator, we can pick any scheme and the results will be unaffected. 

For convenience, let us define
\begin{equation}
\bar{x} = 1-x,\qquad a_s = \frac{g^2}{16\pi^2}, \qquad  
\Theta_0(x) = xm_s^2+\bar{x}m^2-x\bar{x}M^2.
\label{MeffDef}
\end{equation}
In the on-shell scheme we find
\begin{align}
\delta Z^{\text{R}}_\Psi &= -\frac{a_s }{2}\mathfrak{s}^\text{R}+a_s\mathfrak{f}^\text{R}_\Psi \\
& +a_s\int_0^1 dx\ x\qa\log\ta\frac{\Theta_0( x)}{\bar{\mu}^2}\tc-2\bar xM\,\frac{m+xM}{\Theta_0(x)}\qc, \\
\delta Z^{\text{R}}_M &= a_s\ta\frac{1}{2}+\frac{m}{M}\tc \mathfrak{s}^\text{R} + a_s \mathfrak{f}^\text{R}_M\\
& -a_s\int_0^1 dx \ta x+\frac{m}{M}\tc\log\ta\frac{\Theta_0( x)}{\bar{\mu}^2}\tc,
\end{align}
where the singular and finite factors are given by
\begin{equation}\label{Lcounterterms}
\begin{split}
  \mathfrak{s}^\text{R} &= \begin{cases}
\frac{1}{\e} & \text{for R=DR}\\
\log\ta\frac{M_\text{PV}^2}{\bar\mu^2}\tc & \text{for R=PV}
\end{cases}, \\
 \mathfrak{f}^\text{R}_\Psi &= \begin{cases}
0 & \text{for R=DR}\\
\frac{1}{4} & \text{for R=PV}
\end{cases}, \\ \mathfrak{f}^\text{R}_M &= \begin{cases}
0 & \text{for R=DR}\\
-\ta\frac{1}{4} + \frac{m}{M}\tc & \text{for R=PV}
\end{cases}  .
\end{split}
\end{equation}
For DR we defined the typical $\MS$ scale $\bar{\mu}^2 = 4\pi e^{-\gamma_E}\mu^2$. In the case of PV regularization, the scale $\bar{\mu}^2$ is a dummy scale for the sole purpose of having well-defined arguments for the logarithms and, at the same time, isolating the singularity. 

At this point, we already have all the ingredients to deduce the gravitational form factors associated with the proton operator, namely
\[\label{NucleonGFF}
A^\text{R}_\Psi = 1+\delta Z^\text{R}_\Psi, \qquad B_\Psi = 0, \qquad D_\Psi = 0, \qquad \bar{C}_\Psi = 0.
\]
\subsection{Quark vertex}
\label{quark_loop_sec}
To express the results in a somewhat compact form, let us introduce the short-hand notation $\tau^2 = -\Delta^2/4$ and the following definitions
\begin{equation}
\Sigma_0(x,\tau) = \sqrt{\tau^2 + \Theta_0(x)}, \ \ \Lambda_0(x,\tau) = \log\ta\frac{\Sigma_0(x,\tau)+\tau}{\Sigma_0(x,\tau)-\tau}\tc.
\end{equation}
In this way the gravitational form factors associated with the quark operator simply read
\begin{align}
\label{quark_FF_beg}
A^{\text{R}}_q &= \frac{a_s}{6}\mathfrak{s}^\text{R} +a_s \mathfrak{f}^\text{R}_{A}\\
& - a_s\int_0^1 dx\ x\bar{x}\qa \log\ta \frac{\Theta_0(x)}{\bar{\mu}^2} \tc + \frac{\Sigma_0(x;\bar x\tau)}{\bar{x}\tau}\,\Lambda_0(x,\bar x\tau)\qc \notag\\
& \quad +\frac{a_s}{2\tau} \int_0^1 dx \ x\, \frac{\Theta_0(x)+(m+xM)^2}{\Sigma_0(x,\bar x\tau)}\,\Lambda_0(x,\bar x\tau),\notag\\
%%%%%%%%%%%%%%%%%%%%%%%%%%%%%%%%%%%%
B_q &= \frac{a_s M}{\tau} \int_0^1 dx \ x\bar{x}  \,\frac{m+xM}{\Sigma_0(x,\bar x\tau)}\,\Lambda_0(x,\bar x\tau),\\
%%%%%%%%%%%%%%%%%%%%%%%%%%%%%%%%%%%%
D_q &= -\frac{a_s M^2}{\tau^2}\ta\frac{1}{3}+\frac{m}{M}\tc\\
& +\frac{a_sM^2}{\tau^3}\int_0^1 dx\,  \ta x+\frac{m}{M}\tc\Sigma_0(x,\bar x\tau)\, \Lambda_0(x,\bar x\tau), \notag\\
%%%%%%%%%%%%%%%%%%%%%%%%%%%%%%%%%%%%
\bar{C}^{\text{R}}_q &= -\frac{a_s}{2}\ta\frac{1}{3}+\frac{m}{M}\tc\mathfrak{s}^\text{R} + a_s\mathfrak{f}^\text{R}_{\bar{C}} \label{quark_FF_end}\\
& +a_s\int_0^1 dx\ \bar{x}\ta x+\frac{m}{M}\tc\log\ta\frac{\Theta_0(x)}{\bar{\mu}^2}\tc\notag,
\end{align}
where the singular factor $\mathfrak{s}^\text{R}$ is the same as in Eq.~\eqref{Lcounterterms} and the finite factors are here given by
\begin{equation}
\mathfrak{f}^\text{R}_{A} =\begin{cases}
0 & \text{for R=DR}\\
\frac{1}{36} & \text{for R=PV}
\end{cases}, \qquad \mathfrak{f}^\text{R}_{\bar{C}} =\begin{cases}
0 & \text{for DR}\\
\frac{5}{36} + \frac{3m}{4M} & \text{for PV}
\end{cases}.
\end{equation}
We see that, depending on the regularization scheme, the finite part of the gravitational form factors may vary. It appears that the $A_q$ gravitational form factor does not vanish when $\tau\to\infty$. In the case of $\bar{C}_q$, we observe that the whole gravitational form factor is actually $\tau$-independent. Interestingly, the same observations have been made for the electron gravitational form factor at one-loop in QED~\cite{Metz:2021lqv,Freese:2022jlu}. These features can be understood as a reflection of the perturbative nature of the calculations, in particular of the fact that we are considering a pointlike target.

\subsection{Diquark vertex}
\label{diquark_loop_sec}
For the insertion of the EMT operator on the diquark line, everything proceeds in the same way. We obtain 
\begin{align}
A^{\text{R}}_d &=  \frac{a_s}{3}\mathfrak{s}^\text{R}+a_s\ta\frac{2}{3} -10\mathfrak{f}^\text{R}_A\tc-2a_s\int_0^1dx\ x\bar{x}\log\ta\frac{\Theta_0(\bar x)}{\bar{\mu}^2}\tc \notag\\
& \quad+\frac{a_s}{\tau}\int_0^1dx \ x\qa\frac{x M(m+\bar{x}M)}{\Sigma_0(\bar x,\bar x\tau) }-2 \Sigma_0(\bar x,\bar x\tau)  \qc\Lambda_0(\bar x,\bar x\tau),   \\
%%%%%%%%%%%%%%%%%%%%%%%%%%%%%%%%%%%%%%%%
%%%%%%%%%%%%%%%%%%%%%%%%%%%%%%%%%%%%%%%%
B_d &= -\frac{a_sM}{\tau} \int_0^1 dx \ x^2\, \frac{m+\bar{x}M}{\Sigma_0(\bar x;\bar x\tau) }\,\Lambda_0(\bar x,\bar x\tau) , \\
%%%%%%%%%%%%%%%%%%%%%%%%%%%%%%%%%%%%%%%%
D_d & =-\frac{a_sM^2}{\tau^2}\ta \frac{2}{3}+\frac{m}{M}\tc   +\frac{a_sM^2}{\tau^3} \int_0^1 dx \Bigg[ \ta \bar{x}+\frac{m}{M}\tc   \notag\\
&\times\frac{\tau^2(\bar x^2-1)+\Theta_0(\bar x)}{ \Sigma_0(\bar x,\bar x\tau) }  \,\Lambda_0(\bar x,\bar x\tau)\Bigg] , \\
%%%%%%%%%%%%%%%%%%%%%%%%%%%%%%%%%%%%%%%%
\label{CBarD_finalForm}
\bar{C}^{\text{R}}_d &= \frac{a_s}{2}\ta\frac{1}{3}+\frac{m}{M}\tc\mathfrak{s}^\text{R} - a_s \mathfrak{f}^\text{R}_{\bar{C}}\notag\\
& -a_s\int_0^1 dx \ x \ta\bar{x}+\frac{m}{M}\tc \log\ta\frac{\Theta_0(\bar x)}{\bar{\mu}^2}\tc.
\end{align}
Like in the quark sector, $A_d$ does not vanish when $\tau\to\infty$ and $\bar C_d$ is constant.

\section{Renormalization}

It is straightforward to see that all the gravitational form factors, once summed over the proton, quark and diquark contributions, are free from UV divergences. This means that the total symmetric EMT is finite and therefore does not require the introduction of additional counterterms beside the Lagrangian ones. The same observation has been made for an electron state in QED~\cite{Rodini:2020pis}, and is consistent with the general arguments given in Ref.~\cite{Nielsen:1977sy} in the context of non-abelian gauge theories.

\begin{figure}[h]
\centering
\includegraphics[width=0.45\textwidth]{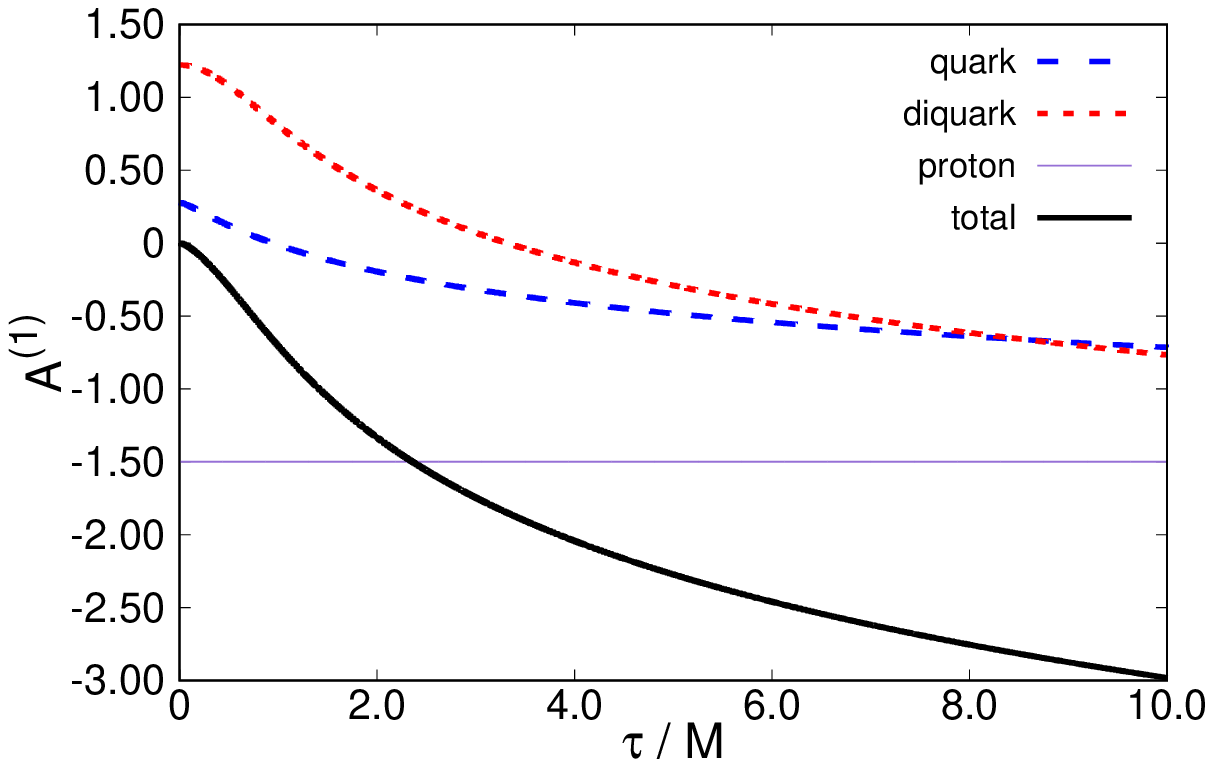}
\quad 
\includegraphics[width=0.45\textwidth]{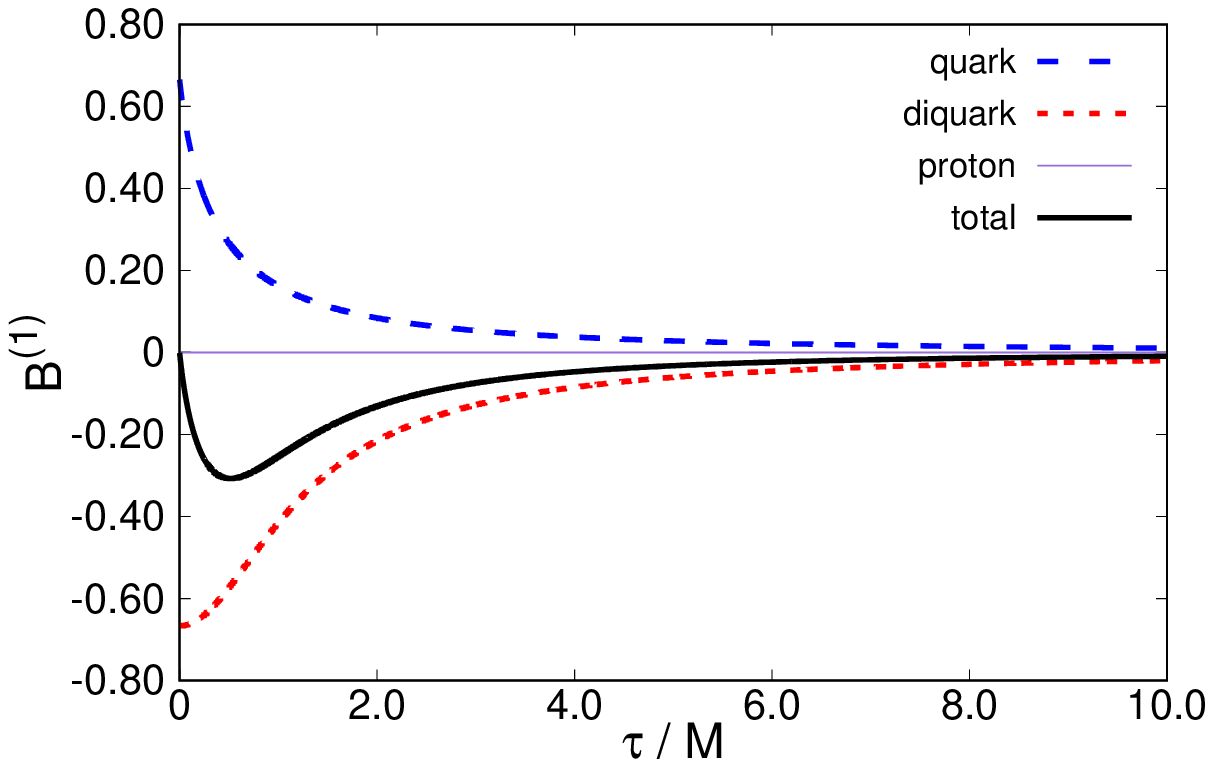}
\\
\includegraphics[width=0.45\textwidth]{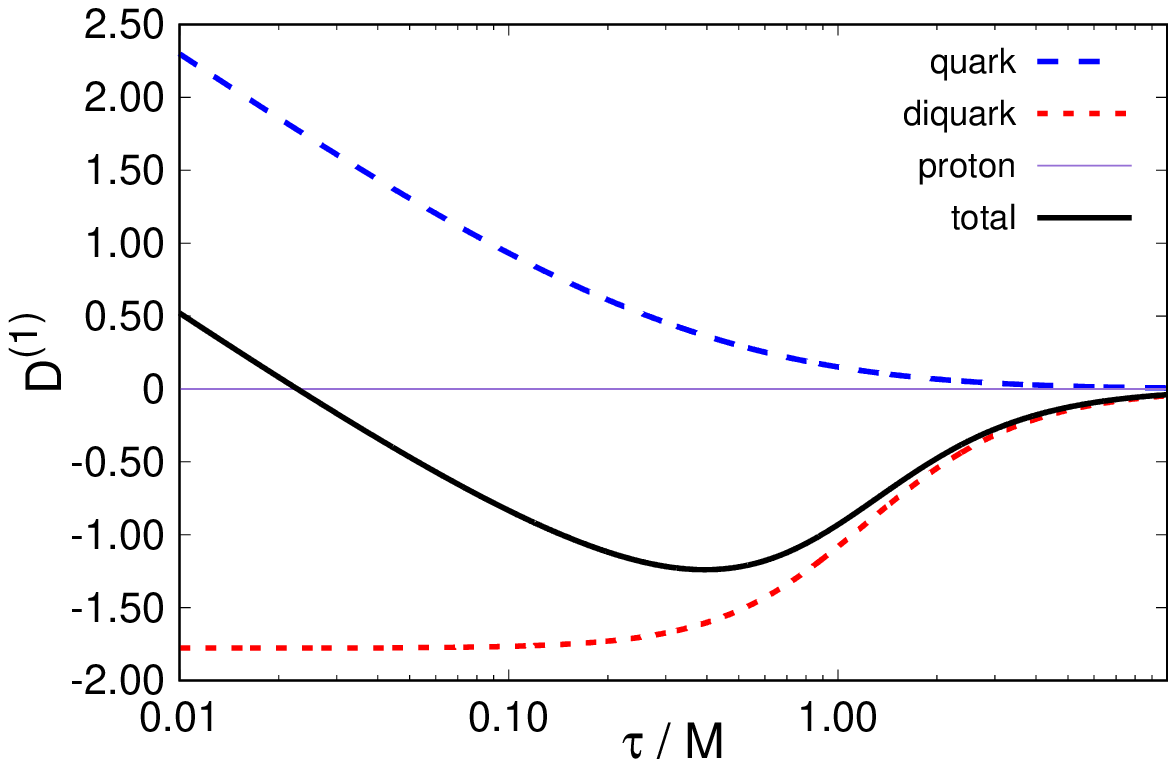}
\caption{One-loop contributions to the gravitational form factors as functions of the dimensionless variable $\tau/M$ for the case $m=0$, $m_s = M$. The thick black solid lines represent the total contributions. For all the panels, the thin solid lines, medium- and short-dashed lines represent the proton, quark and diquark contributions, respectively. All the results are shown in the $\MS$ renormalization scheme and in units of $a_s$. Note that in the case of the $D$ form factor, we used a log scale for the variable $\tau/M$.}
\label{fig:my_label}
\end{figure}

In an MS scheme, limiting ourselves to external proton states, the subtraction of divergences can be performed by trivially removing the singular contribution from the individual proton, quark and diquark gravitational form factors. In Fig.~\ref{fig:my_label} we show the contributions of order $a_s$ to the $A$, $B$ and $D$ gravitational form factors based on the results of the previous section. For simplicity, we chose to illustrate the case $m=0$ and $m_s=M$ in the $\MS$ renormalization scheme.

\section{Sum rules}

A number of constraints on the gravitational form factors can be derived from Poincar\'e symmetry~\cite{Ji:1996ek,Ji:1997pf,Teryaev:1999su,Brodsky:2000ii,Lowdon:2017idv,Cotogno:2019xcl,Lorce:2019sbq}. In particular, four-momentum conservation implies
\begin{equation}\label{momSR}
    \sum_iA_i(0)=1,\qquad \sum_i\bar C_i(\Delta^2)=0,
\end{equation}
and (generalized) angular momentum conservation implies in addition
\begin{equation}\label{AMSR}
    \sum_iB_i(0)=0.
\end{equation}
These contraints should hold at any order in perturbation theory. Let us then write the gravitational form factors as 
\begin{equation}
    X_i=X^{(0)}_i+a_sX^{(1)}_i+\cdots,
\end{equation}
where the upper label indicates the order in $a_s$.

At tree level the Poincar\'e constraints are trivially satisfied since all the gravitational form factors vanish except $A^{(0)}_\Psi(\Delta^2)=1$. Let us now check the $\mathcal O(a_s)$ contributions. Comparing Eqs.~\eqref{quark_FF_end} and~\eqref{CBarD_finalForm}, it is clear after a change of variable $x\mapsto \bar x$ in the integral that
\[
\bar{C}^{(1)}_d(\Delta^2) = -\bar{C}^{(1)}_q(\Delta^2).
\]
Combined with the result $\bar C^{(1)}_\Psi(\Delta^2)=0$ from Eq.~\eqref{NucleonGFF}, we see that the second momentum sum rule in Eq.~\eqref{momSR} is satisfied.

For the gravitational form factors $A_i$, we find in the limit of vanishing momentum transfer
\[
\label{tauToZeroA}
\begin{split}
A^{(1)}_\Psi(0) & =-\frac{1}{2}\,\mathfrak s^\text{R}+ \mathfrak f^\text{R}_\Psi\\
& +\int_0^1dx \ x\qa \log\ta\frac{\Theta_0(x)}{\bar\mu^2}\tc-2\bar{x}M \,\frac{m+xM}{\Theta_0(x)} \qc,\notag\\
A^{(1)}_q(0) & =\frac{1}{6}\ta\mathfrak s^\text{R}-1\tc +\mathfrak f^\text{R}_A\\
& -\int_0^1dx \ x\bar{x}\qa \log\ta\frac{\Theta_0(x)}{\bar\mu^2}\tc-\frac{(m+xM)^2}{\Theta_0(x)}\qc,\notag \\
A^{(1)}_d(0)  & =\frac{1}{3}\,\mathfrak s^\text{R} -10\mathfrak f^\text{R}_A\\
& -2 \int_0^1dx \ x\bar{x}\qa \log\ta\frac{\Theta_0(x)}{\bar\mu^2}\tc-\bar xM\,\frac{m+xM}{\Theta_0(x)}\qc.\notag
\end{split}
\]
Owing to first momentum sum rule in Eq.~\eqref{momSR}, it is expected that the sum of these three contributions should vanish. We find indeed
\[
\begin{split}
&\qa A^{(1)}_\Psi+A^{(1)}_q+A^{(1)}_d\qc(0) = -\frac{1}{6}\\
& +\int_0^1 dx \ \qa (3x^2-2x)\log\ta\frac{\Theta_0(x)}{\bar\mu^2}\tc + x\bar{x}\,\frac{m^2-x^2M^2}{\Theta_0(x)}\qc=0,
\end{split}
\]
where in the last step we integrated by parts and used the relation $\ta1-x\,\frac{d}{d x}\tc\Theta_0(x)=m^2-x^2M^2$.

Finally, since we have
\[
\begin{split}
B^{(1)}_\Psi(0) &=0,\\
 B^{(1)}_q(0) = - B^{(1)}_d(0) &= 2 \int_0^1 dx \ x\bar{x}^2M\,\frac{ (m+x M)}{\Theta_0(x)},
\end{split}
\]
it follows automatically that the (generalized) angular momentum sum rule in Eq.~\eqref{AMSR} is also satisfied.
We notice that to obtain the decomposition of the total angular momentum carried by the quark into its spin and  orbital angular momentum \cite{Ji:1996ek} one would need to include also the anti-symmetric part of the EMT. We leave a detailed discussion for a future work.

\section{$D$-term}

The last gravitational form factor is not constrained by Poincar\'e symmetry. It provides information about the spatial distribution of forces inside the system, and its value at vanishing momentum transfer is known as the $D$-term~\cite{Polyakov:1999gs,Polyakov:2002yz,Polyakov:2018zvc}.

To order $a_s$ in the scalar diquark model, only the quark and diquark sectors contribute to the $D$-term
\[
\begin{split}
D_q(0) &= \frac{2a_s}{3}\int_0^1 dx \ \bar{x}^3M\,\frac{(m+x M)}{\Theta_0(x)} , \\ D_d(0) &= \frac{2a_s}{3}\int_0^1 dx \ x(x^2-3)M\,\frac{(m+x M)}{\Theta_0(x)}.
\end{split}
\]
They are both finite and non-zero. If we assume massless diquark and demand the validity of the stability condition $M<m+m_s$, we find that
\[
D_d(0) \approx \frac{8Ma_s}{3(m-M)}\log(m_s).
\]
Notice that, in line with the common expectation that $D(0)$ is negative for a stable bound state, we observe that $D(0)\to-\infty$ as $m_s\to0$ with $M<m$.
On the other hand, if we assume that $M=m$ and send the scalar diquark mass to zero, we find that the most singular behavior is of the form
\[
D_d(0) \approx \frac{4Ma_s\pi}{3m_s}.
\]
This is identical to the scaling behavior found for an electron state in QED with photon mass regularization~\cite{Metz:2021lqv}. We incidentally note that the situation is different for $D_q(0)$. In fact, in the limit of vanishing quark mass and for $M<m_s$, we find a finite value for $D_q(0)$. In the case $m\to 0$ and $m_s=M$ we find a divergence in $D_q$ (but not in $D_d$, as illustrated by fig.~\ref{fig:my_label}). Specifically, we have
\[
D_q(0) \approx -\frac{2a_s}{3}\log(m).
\]
Let us stress that the two cases $m=M$, $m_s\to 0$ and $m_s=M$, $m\to 0$ present different scaling behaviours: the former is power-like $1/m_s$, the latter is logarithmic $\log(m)$.

The asymptotic behaviour in the large-$\tau$ limit is more cumbersome to extract. For this, let us work with $m_s=M$ and $m=0$. We obtain
\[
\begin{aligned}
D_q &\underset{\tau\gg M}{\approx} -\frac{M^2}{6\tau^2}\qa 2-\log\ta\frac{4\tau^2}{M^2}\tc \qc+\mathcal O\ta\frac{\log\tau^2}{\tau^4}\tc, \\
D_d &\underset{\tau\gg M}{\approx} -\frac{2M^2}{3\tau^2}\qa 1+\log\ta\frac{4\tau^2}{M^2}\tc\qc+\mathcal O\ta\frac{\log\tau^2}{\tau^4}\tc.
\end{aligned}
\]
Changing the values of the masses leads to different numerical factors, but the overall structure remains always
\[
D\sim \frac{c_1 + c_2\log\tau^2}{\tau^2}.
\label{general_large_tau}
\]

\section{Trace}

From the definition of the EMT we can easily see that in DR (hence without ghost fields) the traces of the proton, quark and diquark EMTs are given by
\begin{equation}
\begin{aligned}
    T^\mu_{\Psi\,\mu}&=M\bar\Psi\Psi-\tfrac{1}{2}g\phi\ta\bar\Psi q+\bar q\Psi\tc,\\
    T^\mu_{q\,\mu}&=m\bar qq-\tfrac{1}{2}g\phi\ta\bar\Psi q+\bar q\Psi\tc,\\
    T^\mu_{d\,\mu}&=\partial_\mu \phi\partial^\mu \phi -(2-\epsilon)\ta \partial_\mu \phi\partial^\mu \phi- m_s^2\phi^2\tc.
\end{aligned}
\end{equation}
So, as in QED, the anomaly emerges in DR from the bosonic sector\[
\text{Anomaly} = \e \ta \partial_\mu \phi\partial^\mu \phi - m_s^2\phi^2\tc.
\]
The off-forward matrix elements of the relevant scalar operators are
\begin{align}
&\frac{\braket{M\bar{\Psi}\Psi}}{\bar{u}'u} = M \ta 1+\delta Z_M+\delta Z_\Psi\tc,\\
&\frac{\braket{\frac{1}{2} g \phi\ta\bar{\Psi}q +\bar{q}\Psi\tc}}{\bar{u}'u} = M\, \delta Z_M ,\\
&\frac{\braket{m\bar{q}q}}{\bar{u}'u} =-a_sm\ta\frac{1}{\e}+1\tc +2a_sm \int_0^1 dx\, \bar{x}\log\ta\frac{\Theta_0(x)}{\bar\mu^2}\tc \\
&+ a_sm \int_0^1 dx \Bigg( \frac{(m+xM)^2+\bar{x}^2\tau^2}{2\tau \Sigma_0(x,\bar{x}\tau)} +\frac{3\Sigma_0(x,\bar{x}\tau)}{2\tau}\Bigg)\Lambda_0(x,\bar{x}\tau), \notag\\
&\frac{\braket{\partial_\mu \phi \partial^\mu \phi}}{\bar{u}'u}  = -a_s\ta M+2m\tc\ta\frac{1}{\e}+2\tc\\
& + 2a_s\int_0^1 dx \ \bar x\qa 2(m+\bar xM)-xM\qc\log\ta\frac{\Theta_0(\bar x)}{\bar\mu^2}\tc \notag\\
&+ a_s \int_0^1 dx\Bigg[\frac{(m+\bar{x}M)\ta M^2x^2+\tau^2(x^2-1) \tc}{\tau\Sigma_0(\bar{x},\bar{x}\tau)} \notag\\
& +\frac{\qa 5(m+\bar xM)-2xM\qc\Sigma_0(\bar{x},\bar{x}\tau)}{\tau}\Bigg]\Lambda_0(\bar{x},\bar{x}\tau),\notag\\
&\frac{\braket{m_s^2\phi^2}}{\bar{u}'u} = a_sm_s^2 \int_0^1 dx\ \frac{m+\bar{x}M}{\tau \Sigma_0(\bar{x},\bar{x}\tau)}\,\Lambda_0(\bar{x},\bar{x}\tau).
\end{align}
These results are consistent with the expression for the trace of the general parametrization in Eq.~\eqref{GeneralEMTParametrization_offF}
\begin{equation}
\begin{split}
    \langle T^\mu_{i\,\mu}\rangle& =M\,\bar u' u\Bigg[ A_i+4\bar C_i+\frac{\Delta^2}{4M^2}\ta B_i-3D_i\tc\\
    & +2\epsilon\ta\frac{\Delta^2}{4M^2}\,D_i-\bar C_i\tc\Bigg].
    \end{split}
\end{equation}
In particular, the matrix element of the trace anomaly reads
\[
\braket{\e \ta \partial_\mu \phi\partial^\mu \phi - m_s^2\phi^2\tc} = -a_s\ta M+2m\tc\bar{u}'u 
\]
for $\epsilon\to 0$ and arises purely from the kinetic term.

It has been shown long ago that the forward matrix element of the EMT trace gives the mass of the system~\cite{Shifman:1978zn}
\begin{equation}
    \langle T^\mu_{\phantom{\mu}\mu}\rangle=M\,\bar u u.
\end{equation}
Since the tree-level matrix element of the EMT between proton states already accounts for the total proton mass, we expect that in the limit $\tau\to0$ the trace anomaly is exactly compensated by the $\mathcal O(a_s)$ contribution to the classical expression for the EMT trace:
\[
(T^\mu_{\phantom{\mu}\mu})_\text{class} = \bar{S}\mathbb{M}S+\partial_\mu \phi\partial^\mu \phi -2\ta \partial_\mu \phi\partial^\mu \phi - m_s^2\phi^2\tc.
\label{ClassicalTrace}
\]
This is indeed what is found
\begin{equation}
    \langle (T^\mu_{\phantom{\mu}\mu})_\text{class}\rangle=M\,\bar u u+a_s\ta M+2m\tc\bar{u}u.
\end{equation}
The proton mass can alternatively be obtained from the rest-frame matrix elements of $T^{00}$~\cite{Ji:1994av,Ji:1995sv,Lorce:2017xzd}. Since the four-momentum sum rules are satisfied, we automatically find that $\langle T^{00}\rangle=2M^2$ at rest. Note that the EMT expression~\eqref{EMT_operator} used in our explicit calculations is finite and does not involve any contribution from the trace anomaly, showing that
the latter does not play any intrinsic role in the energy sum rule, in agreement with the analysis of Refs.~\cite{Metz:2020vxd,Lorce:2021xku}.

In PV regularization we find for the trace of the total EMT
\[
\begin{split}
T^\mu_{\phantom{\mu}\mu} &= \bar{S}\mathbb{M}S+\partial_\mu \phi\partial^\mu \phi -2\ta \partial_\mu \phi\partial^\mu \phi - m_s^2\phi^2\tc\\
& -\partial_\mu c\partial^\mu c +2\ta \partial_\mu c\partial^\mu c - M_\text{PV}^2c^2\tc. 
\end{split}
\]
In this case, the anomalous part comes from the ghost sector, even though it is not as transparent as DR. We cannot however isolate the anomalous operator in PV since the ghost sector is also needed to regulate the integrals in the physical sector. This confirms that the anomaly cannot in general be attributed to a particular sector of the theory. Separating the anomaly into contributions associated with different constituents (e.g.~quarks, diquarks, \dots) is therefore a renormalization scheme dependent operation.

\section{Conclusions}

In this work we studied in detail the symmetric energy-momentum tensor of the scalar diquark model to one-loop level in perturbation theory. Contrary to the light-front wave function overlap formalism, the perturbative approach allows us to maintain exact Poincar\'e symmetry throughout the calculations. We extracted the perturbative expressions of all the gravitational form factors using two different regularization methods, namely dimensional and Pauli-villars regularizations. We observed similar pathological behaviors as for the electron in QED, which can be tied to the perturbative nature of our calculations. We checked explicitly that including Lagrangian counterterms are sufficient to make the energy-momentum tensor finite, in agreement with general arguments given in the literature. We also showed that all the Poincar\'e constraints on the gravitational form factors are satisfied. Finally, we demonstrated the consistency of our results with recent discussions about the role played by the trace anomaly in the proton mass.

\appendix
\section{Fourier transform}
\label{Sec_FT}

Fourier transforms of the gravitational form factors can be interpreted in terms of spatial distributions of energy, linear/angular momentum, and forces inside the target~\cite{Polyakov:2002yz,Polyakov:2018zvc,Lorce:2017wkb,Lorce:2018egm,Freese:2021czn}. Three- and two-dimensional Fourier transforms are respectively defined as
\begin{align}
\hat{F}(r) &= \int \frac{d^3\Delta}{(2\pi)^3}\,e^{-i\bm{\Delta}\cdot \bm{r}} F(-\bm{\Delta}^2) \\
& = \int_0^\infty \frac{d\kappa}{2\pi^2}\  \kappa^2 F(-\kappa^2) j_0(\kappa r),\notag\\
\tilde{F}(b_\perp)&= \int \frac{d^2\Delta_\perp}{(2\pi)^2}\,e^{-i\bm{\Delta}_\perp\cdot \bm{b}_\perp} F(-\bm{\Delta}^2_\perp) \\
& =  \int_0^\infty \frac{d\kappa}{2\pi}\  \kappa F(-\kappa^2) J_0(\kappa b_\perp),\notag
\end{align}
where $j_0$ and $J_0$ are the spherical and cylindrical Bessel functions of the first kind. Any constant term in the gravitational form factors has singular Fourier transformation, contributing as $\delta^3(\bm{r})$ or $\delta^2(\bm{b}_\perp)$. We will discard any such contributions in the following discussion, since they emerge as pathological features of the perturbative nature of the presented results.

For the $B$ form factors we find in two and three dimensions
\begin{align}
\begin{pmatrix}\tilde{B}_q\\ \hat B_q\end{pmatrix} &= \frac{2 a_s M}{\pi} \int_0^1 dx \ x(m+xM) \begin{pmatrix}K_0^2(\zeta_q)\\ \frac{1}{r}K_0(2\zeta_q)\end{pmatrix}, \\
\begin{pmatrix}\tilde{B}_d\\ \hat B_d\end{pmatrix} &= -\frac{2 a_s M}{\pi} \int_0^1 dx \ \frac{x^2}{\bar{x}}(m+\bar{x}M) \begin{pmatrix}K_0^2(\zeta_d)\\ \frac{1}{r}K_0(2\zeta_d)\end{pmatrix}, 
\end{align}
where we defined
\[
\zeta_q = \frac{\sqrt{\Theta_0(x)}}{\bar{x}}\begin{pmatrix}
    b_\perp \\
    r
\end{pmatrix}, \qquad \zeta_d = \frac{\sqrt{\Theta_0(\bar x)}}{\bar{x}}\begin{pmatrix}
    b_\perp \\
    r
\end{pmatrix}.
\]

For the $A$ form factors, we isolate and remove the constant contribution (in the momentum transfer) that leads to singular Fourier transforms
\begin{align}
A_q^{\text{R,const}} &=\frac{a_s}{6}\ta\mathfrak{s}^\text{R}-1\tc+a_s\mathfrak f^\text{R}_A - a_s \int_0^1 dx \ x\bar{x}\log\ta\frac{\Theta_0(x)}{\bar\mu^2}\tc, \\
A_d^{\text{R,const}} &= \frac{a_s}{3}\mathfrak{s}^\text{R}-10\mathfrak f^\text{R}_A - 2a_s \int_0^1 dx \ x\bar{x}\log\ta\frac{\Theta_0(x)}{\bar\mu^2}\tc.
\end{align}
Subtracting these constant terms, we find 
\begin{align}
\begin{pmatrix}\tilde{A}_q \\ \hat{A}_q \end{pmatrix}&= \frac{a_s}{\pi} \int_0^1 dx \ \frac{x}{\bar{x}} \Bigg[ (m+xM)^2 \begin{pmatrix} K_0^2(\zeta_q)\\ \frac{1}{r}K_0(2\zeta_q)\end{pmatrix} \notag\\
& + \Theta_0(x) \begin{pmatrix} K_1^2(\zeta_q)\\ \frac{1}{r}K_2(2\zeta_q)\end{pmatrix} \Bigg], \\
%%%%%%%%%%%%%%%%%%%%%%%%%%%%%%%%%%%%%%%%%%%%%
\begin{pmatrix}\tilde{A}_d \\ \hat{A}_d \end{pmatrix} &= \frac{2a_s}{\pi} \int_0^1 dx \ \frac{x}{\bar{x}} \Bigg[ xM(m+\bar{x}M) \begin{pmatrix} K_0^2(\zeta_d)\\ \frac{1}{r}K_0(2\zeta_d)\end{pmatrix} \notag\\
& +\Theta_0(\bar x) \begin{pmatrix} K_1^2(\zeta_d)-K_0^2(\zeta_d)\\ \frac{1}{r}K_2(2\zeta_d)-\frac{1}{r}K_0(2\zeta_d)\end{pmatrix}\Bigg].
\end{align}

For the $D$ form factors we find
\begin{align}
\begin{pmatrix}\tilde{D}_q\\ \hat{D}_q\end{pmatrix} &= \frac{4a_s M}{\pi}\int_0^1 dx \ \bar{x}(m+x M) \\
& \times \int_0^1 dz \ \frac{\sqrt{1-z^2}}{z}\begin{pmatrix}K_0\ta \frac{2}{z}\zeta_q\tc\\ \frac{1}{2r}\exp\ta-\frac{2}{z}\zeta_q\tc\end{pmatrix}, \\
%%%%%%%%%%%%%%%%%%%%%%%%%%%%%%%%%%%%%%%%%%%%%%%%%%%
\begin{pmatrix}\tilde{D}_d\\ \hat{D}_d\end{pmatrix} &= -\frac{4a_s M}{\pi}\int_0^1 dx \ \frac{m+\bar{x} M}{\bar{x}} \\
& \times \int_0^1 dz \ \frac{1-\bar{x}^2(1- z^2)}{z\sqrt{1-z^2}}\begin{pmatrix}K_0\ta \frac{2}{z}\zeta_d\tc\\ \frac{1}{2r}\exp\ta-\frac{2}{z}\zeta_d\tc\end{pmatrix}.
\end{align}
The integral
\[
F(y) = \int_0^1 dz \ \frac{\sqrt{1-z^2}}{z}\begin{pmatrix}K_0\ta \frac{y}{z}\tc\\ \exp\ta-\frac{y}{z}\tc\end{pmatrix}
\]
is a solution of the differential equation
\[
\ta 1-y\,\frac{d}{d y}\tc F(y) = \begin{pmatrix}
    \frac{1}{2}K_0^2\ta \frac{y}{2}\tc \\ 
    K_0\ta y\tc
\end{pmatrix}.
\]

 To study the physics in position space, we introduce the tangential and radial pressures $p_{t,r}$ in three dimensions and $\sigma_{t,r}$ in two dimensions. We also introduce the energy densities $\varepsilon$ in three dimensions and $\rho$ in two dimensions. The definitions in terms of the gravitational form factors are~\cite{Lorce:2018egm}
\[
\begin{split}
p_t(r)/M & = \frac{1}{4M^2r}\frac{d}{d r}\ta r\, \frac{d}{d r} \hat{D}(r)\tc, \\
p_r(r)/M  &= \frac{1}{2M^2r}\frac{d}{d r} \hat{D}(r), \\
\varepsilon(r)/M &= \hat{A}(r) + \frac{1}{4M^2r^2}\frac{d}{d r} \qa r^2 \frac{d}{d r}\ta\hat{B}(r)-\hat{D}(r)\tc\qc, \\ 
\sigma_t(b_\perp)/M  &= \frac{1}{4M^2}\frac{d^2}{d b^2_\perp}\tilde{D}(b_\perp), \\
\sigma_r(b_\perp)/M  &= \frac{1}{4M^2b_\perp}\frac{d}{d b_\perp}\tilde{D}(b_\perp), \\
\rho(b_\perp)/M &= \tilde{A}(b_\perp) \\
& + \frac{1}{4M^2b_\perp^2}\frac{d}{d b_\perp} \qa b_\perp^2 \frac{d}{d b_\perp}\ta\tilde{B}(b_\perp)-\tilde{D}(b_\perp)\tc\qc.
\end{split}
\]
The $\bar C$ form factors being constant, their contributions have been discarded in the above expressions. We stress that the $D$ form factors have non-singular Fourier transforms $\tilde{D}$ and $\hat{D}$, but their derivatives present singular behaviors. Since derivatives in the radial variable of order $n$ are related to Fourier transforms of $\kappa^{n} D(-\kappa^2)$, it is trivial to conclude that the fall-off of $D(-\kappa^2)$ for large values of $\kappa=2\tau$ given in Eq.~\eqref{general_large_tau} is not fast enough to guarantee the absence of singular contributions to pressure and shear in two and three dimensions.
The singular contributions are however fundamental to ensure that the von Laue condition for mechanical equilibrium~\cite{Laue:1911lrk} 
\[
\int_0^\infty d r \ r^2 (p_r+2p_t) = 0
\]
is satisfied. Interestingly, the combination of singular and regular contributions resemble the definition of $+$-distributions commonly used in QCD
\begin{equation}
    \qa\frac{f(z)}{1-z}\qc_+\equiv\frac{f(z)}{1-z}-\delta(1-z)\int_0^z dy\,\frac{f(y)}{1-y}.
\end{equation}
Similar considerations apply to the derivatives of the $B$ form factor. 
We notice that usually the validity of the von Laue condition is viewed as the result of a compensation between regions of positive and negative pressures inside the system. In this case the size of one of the region shrinks to zero, becoming the singular contribution that still ensures the validity of the stability condition.
For this reason, while the perturbative approach ensures that Poincar\'e symmetry is preserved, which is the focus of this work, one cannot consider that the one-loop results for the gravitational form factors provide a realistic picture of a bound state, and even less of the proton structure.

\bibliographystyle{JHEP}
%\bibliography{biblio}

\providecommand{\href}[2]{#2}\begingroup\raggedright\endgroup


\end{document}